\documentclass[]{raa}            
\usepackage{graphicx,times}
\usepackage{natbib}

\usepackage{amsmath}
\usepackage{amssymb}
\usepackage{multirow}
\usepackage{xcolor}

\providecommand{\bm}[1]{\mbox{\boldmath$#1$\unboldmath}}

\def\etal{et al.}

\begin{document}

   \title{Merging Strangeon Stars
}

 \volnopage{ {\bf 2017} Vol.\ {\bf X} No. {\bf XX}, 000--000}
   \setcounter{page}{1}

   \author{Xiao-Yu Lai
      \inst{1,2}
   \and Yun-Wei Yu
      \inst{3}
   \and En-ping Zhou
      \inst{4,5}
    \and Yun-Yang Li
      \inst{4}
   \and Ren-Xin Xu
      \inst{4,5}
   }

   \institute{School of Physics and Mechanical \& Electrical Engineering, Hubei University of Education, Wuhan 430205, China; {\it laixy@pku.edu.cn}\\
        \and
             Xinjiang Astronomical Observatory, Chinese Academy of Science, Urumqi 830011, China
        \and
             Institute of Astrophysics, Central China Normal University, Wuhan 430079, China
        \and
             School of Physics, Peking University, Beijing 100871, China
         \and
             Kavli Institute for Astronomy and Astrophysics, Peking University, Beijing 100871, China\\
\vs \no
   {\small Received [year] [month] [day]; accepted [year] [month] [day]}
}

\abstract{
The state of supranuclear matter in compact star remains puzzling, and it is argued that pulsars could be strangeon stars.
What if binary strangeon stars merge?
This kind of merger could result in the formation of a hyper-massive strangeon star, accompanied by bursts of gravitational waves and electromagnetic radiation (and even strangeon kilonova explained in the paper).
The tidal polarizability of binary strangeon stars is different from that of binary neutron stars, because a strangeon star is self-bound on surface by fundamental strong force while a neutron star by the gravity, and their equations of state are different.
Our calculation shows that the tidal polarizability of merging binary strangeon stars is favored by GW170817.
Three kinds of kilonovae (i.e., of neutron, quark and strangeon) are discussed, and the light curve of the kilonova AT 2017gfo following GW170817 could be explained by considering the decaying strangeon nuggets and remnant star spin-down. 
Additionally, the energy ejected to the fireball around the nascent remnant strangeon star, being manifested as a Gamma-ray burst (GRB), is calculated.
It is found that, after a promote burst, an X-ray plateau could follow in a timescale of $10^{2-3}$ s.
Certainly, the results could be tested also by further observational synergies between gravitational wave detectors (e.g., aLIGO) and X-ray telescopes (e.g., Chinese HXMT and eXTP), and especially if the detected gravitational wave form is checked by peculiar equation of state provided by the numerical relativistical simulation.
}

   \authorrunning{Lai, Yu, Zhou, Li \& Xu}            
   \titlerunning{Merging Strangeon Stars}  
   \maketitle

\section{Introduction}

The nature of pulsar-like compact stars is essentially a central question of the fundamental strong interaction (explained in quantum chromo-dynamics) at low energy scale, the solution of which still remains a challenge though tremendous efforts have been tried.
This kind of compact objects could actually be strange {\it quark} stars instead of neutron stars, if strange quark matter in bulk may constitute the true ground state of the strong-interaction matter rather than $^{56}$Fe (the so-called Witten's conjecture~\citep{Witten1984}).

From astrophysical points of view, however, it is proposed that strange cluster matter could be absolutely stable and thus those compact stars could be strange {\it cluster} stars in fact.
This proposal could be regarded as a {\it general Witten's conjecture}: strange matter in bulk could be absolutely stable, in which quarks are either free (for strange quark matter) or localized (for strange cluster matter).
Strange cluster with three-light-flavor symmetry is renamed ``strangeon'', being coined by combining ``strange nucleon'' for the sake of simplicity.
A strangeon star can then be thought as a 3-flavored gigantic nucleus, and strangeons are its constituent as an analogy of nucleons which are the constituent of a normal (micro) nucleus.
However, the most important issue is to find observational evidence to verify or disaffirm the proposal.

The observational consequences of strangeon stars show that different manifestations of pulsar-like compact stars could be understood in the regime of strangeon stars (see the review by~\cite{LX2017} and references therein).
Since it could be possible that pulsar-like compact stars are actually strangeon stars, neutron star binaries could actually be strangeon star binaries.
The coalescence of strangeon stars in a binary will release signals of gravitational waves as well as electromagnetic radiation, both of which would be detected.
These signals, other than those from isolated stars, would provide additional useful ways to constrain the properties of pulsar-like compact stars.
In this paper, we focus on the possible different electromagnetic behaviors of the merger strangeon stars, and the energy ejection to the fireball of newborn remnant after the merge.

During the phase of tidal disruption of the stars as they approach each other before the final merge, a small fraction of the total mass of both stars should be released.
Although strangeon stars could be in a solid state at low temperature ($\lesssim 1$ MeV), during the phase of tidal disruption and coalescence the temperature of both stars in the binary would rise so that the strangeon stars would be phase-converted to a liquid state.
For a binary of two strangeon stars both with the typical mass $\sim 1.4 M_\odot$, the remnant could be a hyper-massive strangeon stars with mass $\sim 2.6 M_\odot$, which would still be in a liquid state until it cools down to $\sim 1$ MeV.
In this case, strangeon stars in a binary just before merger and the remaining strangeon star at the early stage would behave more or less like conventional quark stars.

Hydrodynamical simulations of the coalescence of quark stars have been performed~\citep{Bauswein2010}, under the equation of state within the MIT bag model.
Different from the case of neutron star merger which forms dilute halo structures around the remnant, the merger of quark stars results in a clumpy strange matter disk.
For a binary of two quark stars with the equal mass $\sim 1.35 M_\odot$, simulations show~\citep{Bauswein2010} that the mass of the disk around the remnant  is about $0.1 M_\odot$ and the mass of the ejecta is about $10^{-3} M_\odot$.
The ejected small lumps of strange matter are called strangelets, and large lumps of strange matter are called strange nuggets, both of which would present in cosmic rays~\citep{Madsen2005}.
Interestingly, ~\cite{Geng2015} studied the coalescence of quark planets with quark stars and show that it could be a new kind of gravitational wave sources.
Merger of strangeon stars in a binary has not been calculated yet, but we could infer that the mass of ejecta could be larger than that in the case of binary quark stars, as the binding of strangeons in strangeon stars should be weaker than that of quarks in quark stars.   
Therefore, although strangeon stars in a binary just before merger and the remnant strangeon star at the early stage would behave like conventional quark stars, the mass of the ejecta and the disk around the remnant would be larger than that of a conventional quark star binary.

Some simulations of binary neutron star merges, however, show that the released mass could be from $10^{-4} M_\odot$ to $10^{-2} M_\odot$~\citep{Goriely2011,Piran2013}, depending on parameters such as the stiffness of the equation of state, the total mass of the binary and the production of the merge.
After thermalization, the ejecta would play an important role in the afterglows, such as the radiation of optical and near infrared~\citep{LP1998}.
If the total mass of ejecta is about $10^{-2} M_\odot$, the transient event is called ``kilonova''~\citep{Metzger2010}.
If an massive millisecond magnetar is formed after the merge, the transient event is called ``merger-nova''~\citep{YGZ2013}.
However, if the compact star binaries are actually strangeon star binaries, the ejecta from tidally elongated strangeon stars and their merger could be relatively less than that from binary normal neutron star merging, because a strangeon star is self-bound on surface by fundamental strong force while a neutron star by the gravity.
And then what kind of electromagnetic radiation would be emitted after merging?

The gravitational event GW170817 and its mutiwavelength electromagnetic counterparts open a new era that the nature of pulsar-like compact stars could be tested crucially.
Unlike the neutron-rich ejecta in the case of merging neutron stars that will decay and radiate, the ejecta composed of strangeon nuggets would not lead to r-process nucleosynthesis.
However, merging strangeon stars could also lead to ``kilonova'' by decay of strangeon nuggets and the spindown of the remanent compact star.
The observed ``blue component'' of the kilonova AT 2017gfo following GW170817 could be powered by the decay of ejected strangeon nuggets, while the late ``red component'' could be powered by the spin-down of the remnant strangeon star after mergering.
On the other hand, the clumpy strange matter disk instead of a dilute halo structure around the remnant would make it possible to detect the thermal radiation of the remnant shortly after the merger.
In this paper, we also consider the energy ejection via the cooling process of the remnant strangeon star, which could be tested by observations, e.g., by HXMT and the future eXTP.

This paper is arranged as follows. 
In Section 2 we test the strangeon star model with the constraints on tidal polarizability by GW170817.
In Section 3 we introduce the ``kilonova'' by decay of strangeon nuggets, which are called ``strangeon kilonova''. 
We derive the bolometric light curve of strangeon kilonova, which shows that under reasonable parameters, the light curve of the kilonova AT 2017gfo could be fitted by considering the decaying strangeon nuggets and remnant star spin-down. 
The thermal radiation of the remnant massive strangeon star are calculated in Section 4.
Summary and discussions are made in Section 5.

\section{Tidal polarizability tested by GW170817}

Recently, a gravitational wave event with companion mass of $\sim
1.4M_\odot$ is discovered~\citep{ligo2017}.
It is found that the tidal polarizability (see Eq.(\ref{Lambda}) below)
of individual companion star would not be larger than $10^3$, and
some of relatively soft state equation of normal neutron star (e.g.,
SLy and APR4) could be favored~\citep{EoS2009}, with maximum masses of $M_{\rm max}\sim 2.05M_\odot$ for SLy and $M_{\rm max}\sim 2.21M_\odot$ for APR4 of \{$npe\mu$\}-matter, although the hyperon puzzle is unavoidable in nucleon star models~\citep{Bombaci2017}.
Nevertheless, for strangeon star with mass $\sim 1.4 M_\odot$, the
radius could be smaller than that of APR4, even though the equation
of state is still very stiff so that the maximum mass would reach
$\sim 3M_\odot$.
Clear evidence for strangeon star could be obtained if a massive pulsar as high as $\sim
2.3M_\odot$ is discovered by advanced facilities (e.g., Chinese FAST).

\begin{figure}[h!!!]
	\centering
	\includegraphics[width=9.0cm, angle=0]{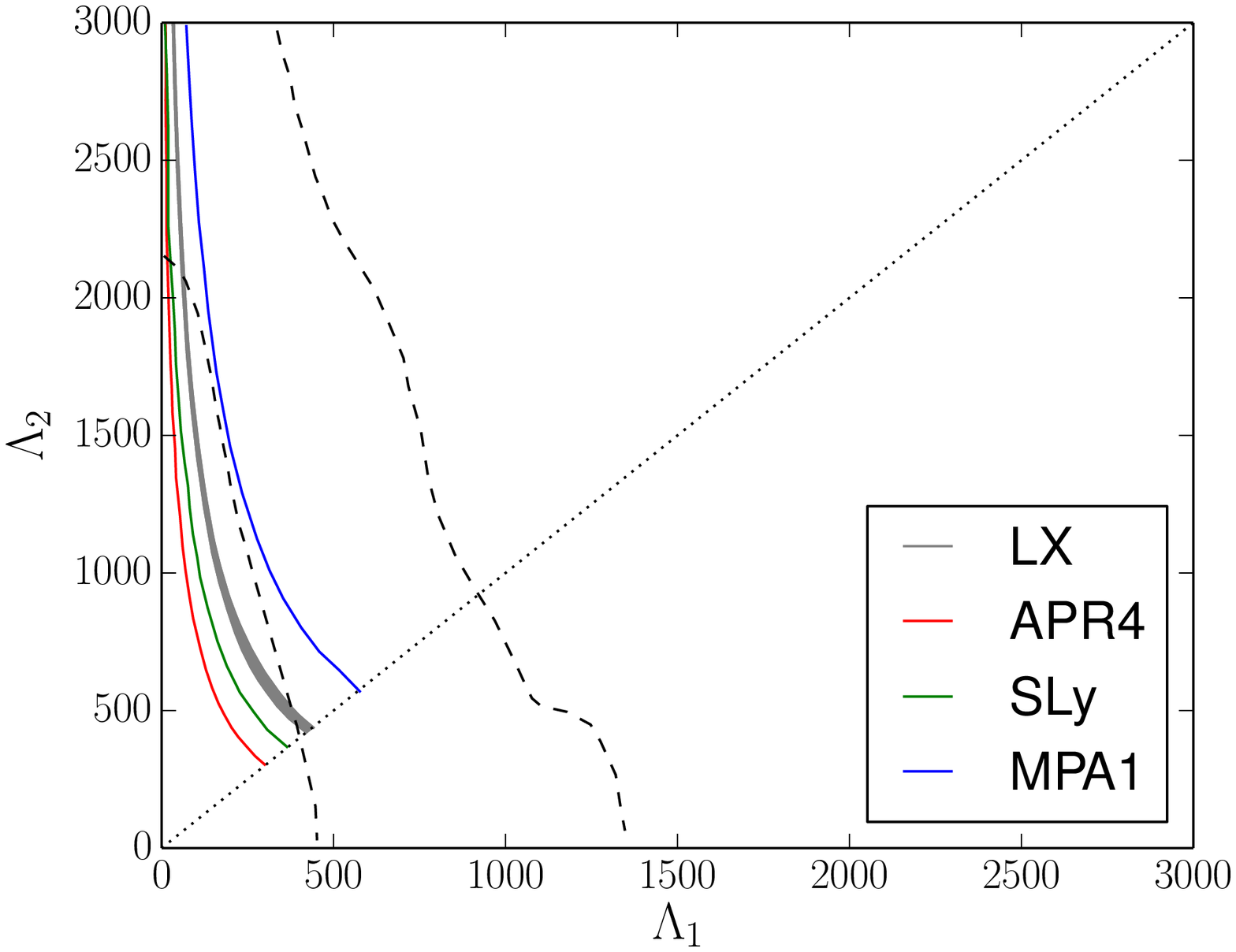}
	\includegraphics[width=9.0cm, angle=0]{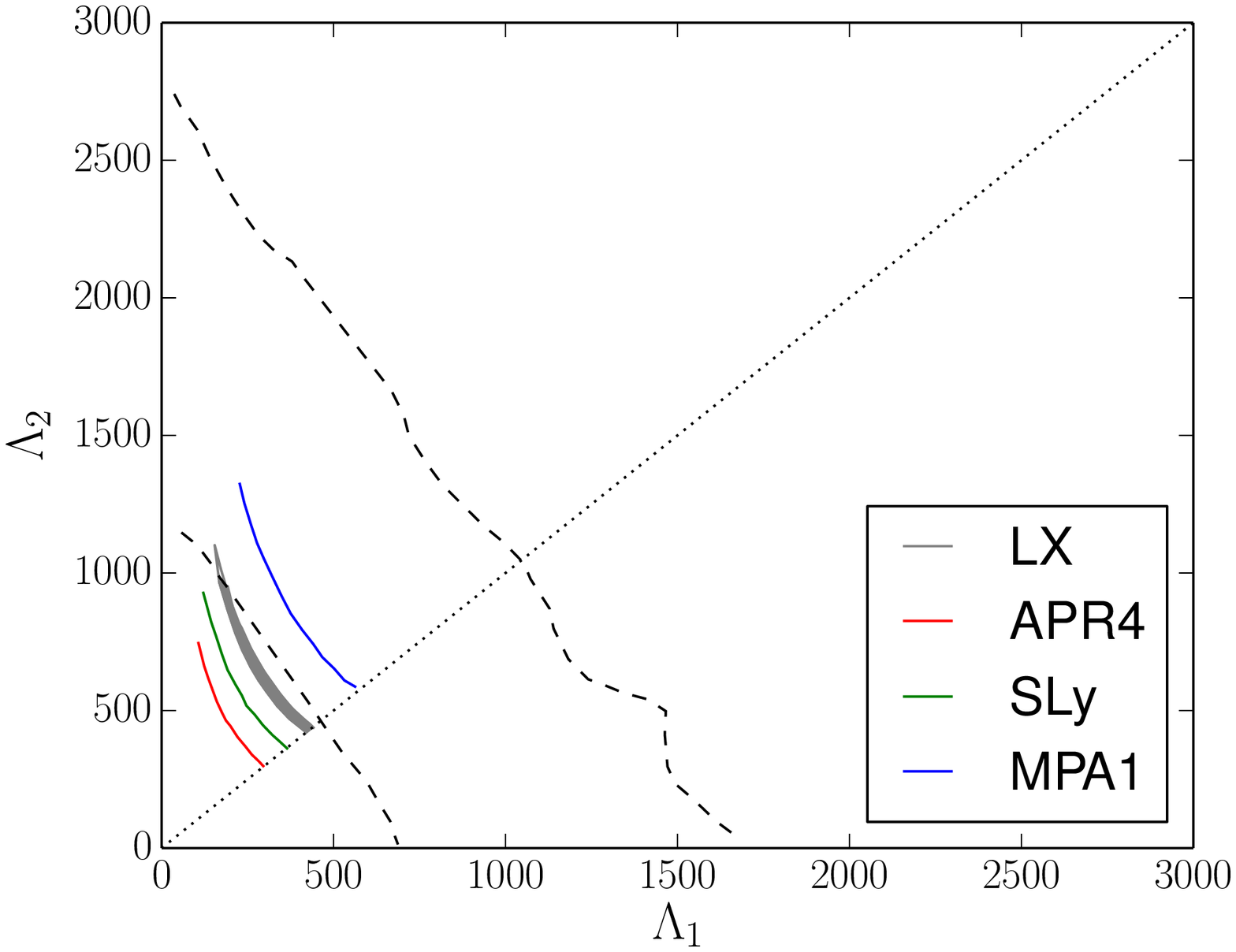}
	\begin{minipage}[]{85mm}
		
		\caption{Comparison of the tidal polarizability of strangeon star model
			with the results obtained by~\cite{ligo2017}. The upper panel is for the higher spin prior (i.e., $|\chi|\leq0.89$) and the lower panel is for the lower spin prior (i.e., $|\chi|\leq0.05$). The dotted diagonal line indicates the boundary of $\Lambda_1=\Lambda_2$. The dashed lines are 50\% (the one to the bottom left corner) and 90\% posterior (the one to the top right) distribution contour for independent 
			$\Lambda_1$ and $\Lambda_2$ priors with a post-Newtonian waveform~\citep{ligo2017}. As can be seen,
			the strangeon star EOS (labeled as 'LX') is favored as the other two soft EOSs (APR4 and SLy) in both cases. }\end{minipage}
	\label{fig:tidal}
\end{figure}

Similar as BNS mergers, mass quadrupole moment will be induced in the late
inspiral phase of a binary strangeon star merger, due to the tidal field of each companion on the other. This property can be characterized by the following relationship,
\begin{equation}
Q_{ij}=\lambda(m)\mathcal{E}_{ij},
\end{equation} 
where the tensor $\mathcal{E}_{ij}$ is an external tidal field and $Q_{ij}$ is
the induce mass quadrupole moment. In this relationship, $\lambda(m)$ is 
a function of the stellar mass which also depends on internal structure of 
the star and related to the so called 
$l=2$ tidal love number $k_2$ by
\begin{equation}
k_2=\frac{3}{2}\lambda R^{-5}.
\end{equation}
The induced mass quadrupole moment will accelerate the coalescence, hence can be constrained
by GW observations \citep{FlanaganHinderer08}. 

In order to test the strangeon star model with the constraints on tidal polarizability by GW170817 and future observations,
we have calculated $k_2$ for strangeon star EOS by introducing a static $l=2$ perturbation to 
the TOV solution \citep{Hinderer2008}. Additionally, the finite surface energy density of the strangeon star requires a special treatment on the boundary condition to obtain the 
correct result \citep{Postnikov2010}. Once we have $k_2$ calculated, 
it's straightforward to obtain the dimensionless tidal polarizability by
\begin{equation}
\Lambda=\lambda/M^5=\frac{2}{3}k_2(R/M)^5. \label{Lambda}
\end{equation}

According to the observation of GW170817, the dimensionless tidal polarizability of a star with 1.4$\,M_\odot$ (i.e., $\Lambda(1.4)$) is constrained with an upper limit of 800 for low spin case and 1400 for high spin case \citep{ligo2017}. The corresponding value for strangeon star EOS is 381.9 in our calculation. This result indicates that although the strangeon star EOS is so stiff that the TOV maximum mass would reach $\sim3M_{\odot}$, the tidal polarizability is actually similar to those soft EOS models such as APR4 and SLy which are favored by GW170817.

For a more systematic test, we have employed the 90\% most probable fraction of component masses $m_1$ and $m_2$ for GW170817, to calculate $\Lambda_1$ and $\Lambda_2$ with strangeon star EOS. The result is compared with the posterior distribution for $\Lambda_1$ and $\Lambda_2$ with post-Newtonian waveforms as well as three other neutron star EOSs obtained by~\cite{ligo2017}. As shown in Fig.1, in both high spin and low spin cases, strangeon star EOS is favored by the constraints. Particularly, considering that the strangeon star will be in a solid state until in the very late inspiral when the tidal heating might lead to a phase transition of the star to liquid state, the actual contribution from the mass quadrupole moment to the waveform  will be even less significant. In this case, the strangeon star model is actually more favored by GW170817.

\section{Strangeon matter kilonova}

After discussing the tidal polarizability, we turn to the tidal destruction and head-on collision of merging compact stars.
Certainly, the ejecta and radiation features depend on the composition of stellar material.
For normal neutron stars in which neutron fluid is dominant, a neutron-rich environment forms, resulting in r-process nucleosynthesis as well as Li-Paczy{\'n}ski nova~\citep{LP1998}, termed ``{\it neutron matter}'' kilonova afterwards \citep{Metzger2010}.
However, it is also argued that Big-Bang like nucleosynthesis could occur for strange quark star merger, reaching the Fe peak but not the lanthanides and gold~\citep{sqm}. We may then call its observational consequence as ``{\it quark matter}'' kilonova.
Will the tidal and collision of two strangeon stars behave similarly? Unfortunately, this process has never been studied extensively yet, but nonetheless there could be different physics from both scenarios above, except the note that r-process nucleosynthesis would also not occur during the coalescence of strangeon stars~\citep{xu2015}.
We are introducing these three kinds of kilonova here, focusing on similarities and differences among them.

{\it Neutron kilonova}.
\cite{LP1998} put forth the idea that merging neutron stars could produce neutron-rich ejecta in which r-process nucleosynthesis will happen. 
The neutron-rich material can be ejected at a speed of about $0.1c$ due to the tidal polarization and probably maintains a low temperature, i.e., a large neutron fraction. 
Alternatively, the ejecta might also originate from the outflows of the accretion disk with a speed of $0.1-0.2c$ and could be proton rich \citep[e.g.,][]{Barzilay2008}.
As the ejecta expands as an envelop, a rapid electro-magnetic transient--the kilonova--powered by $\beta-$ decay and fission would occur.
\cite{Metzger2010} further develop the model and find that the radioactive heating rate of the kilonova robustly peaks on a timescale $t_\mathrm{peak} \sim 1$ day which results in a light curve with similar $t_\mathrm{peak}$. 
This distinguishes kilonova from the Type Ia supernova which has $t_\mathrm{peak} \sim$ weeks since the latter has a larger amount of ejecta and the fuel ($^{56}\mathrm{Ni}$) with a longer half-life. 
The r-process would also produce a significant amount of lanthanide elements that is optically thick in the UV bands due to the line blanketing effect and lead to a reddened spectrum. 
To a great extent, the optical follow-up of GW170817 confirms the prediction of the kilonova scenario. 
Specifically, the early-time spectra reveal a relativistic expanding ($\sim0.2c$, i.e., rapidly cooling) photosphere with a blackbody temperature of $10^4 \mathrm{K}$. 
As the transient fades out in subsequent days, the spectral peak moves redward to the infrared end \citep{SBscience}. 
However, a fast-fading blue component dominating the UV/optical bands in the early spectra was not expected from the model \citep[e.g.,][]{SBscience,DMscience,CPapjl}. 
This can be attributed to a lanthanide-poor ejecta that originate from the squeezing of the neutron stars or the post-merger disk wind \citep[e.g.,][]{MBapjl,KDnature}, although alternatives are also proposed \citep[e.g.,][]{Piro2017, Ioka2017}. 
Even by assuming two components of the ejecta with different masses, velocities and lanthanide fractions, the modeling of the long term spectral evolution is not completely satisfying \citep{KCscience,CRapjl}.
Therefore, even if the kilonova is the true nature beneath the optical transient of GW170817, there is still a long way to go to figure out the nuclear-physical, dynamical and radiative details of the merger event.

{\it Quark kilonova}.
It is shown by \cite{sqm14} that no significant strangelet would survive in the ejecta after the merger and most of the ejecta would decay into protons and neutrons.
Therefore, heavy elements would be built in a bottom-up manner in analogy to the Big-Bang nucleosynthesis.
Assuming a subrelativistic free expansion speed of about $0.2c$, \cite{sqm} find that the proton-neutron equilibrium will freeze out on a millisecond timescale, resulting in a final neutron-to-proton ratio of about $0.7-0.8$, which is significantly higher than that from the Big-Bang nucleosynthesis but lower than that for the r-process element synthesis. 
Consequently, nucleosynthesis stops at the Fe-peak elements and a total absence of lanthanides is expected.
Nevertheless, many radio-active low-mass elements (with mass number $<70$) being produced power a light curve peaking on a one-day timescale. 
The luminosity of this strange quark kilonova drops by 2 orders of magnitude in about a week, which is consistent with the optical/IR counterpart associate with GW170817 \citep{KCscience,SMapjl}. 
However, a detailed match of the spectral evolution is needed before drawing further conclusions.

{\it Strangeon kilonova}.
Merging of strangeon stars has not been studied yet, but we could qualitatively describe some possible consequences which could be compared to observations.
Being heated by the tidal process, two strangeon stars in a binary before coalescence would be phase-converted to a liquid state, so they would behave like conventional quark stars.
Binding of strangeons in strangeon stars (which are bound by residual chromo interaction) should be weaker than that of quarks in quark stars (which are bound by chromo interaction), we could infer that the mass of ejecta of binary strangeon stars could be larger than that of binary quark stars.   
Therefore, if the mass of the ejecta of merging binary quark stars is about $10^{-3}M_\odot$~\citep{Bauswein2010}, then we could assume that the mass of the ejecta of merging strangeon stars could be as high as $10^{-2}M_{\odot}$. 
Although strangeon matter in bulk could be more stable than nuclear matter, the ejected strangeon nuggets (small lumps of strangeon matter) could be unstable under the strong and weak interactions.
We could put the lower limit of critical baryon number $A_c$ of stable strangeon nuggets to be $10^9-10^{10}$, corresponding to the strangeon nuggets with size comparable to the Compton wavelength of electrons~\citep{LX2017}, and strangeon nuggets with baryon number $A<A_c$ are unstable and will decay to protons and neutrons.
The the luminosity of the decay is
\begin{equation}
L_{\rm strangeon\ kilonova} \sim 10^{42} {\rm erg\ s^{-1}} \left(\frac{M_{\rm unstable}}{10^{-4}M_\odot}\right) \left(\frac{\Delta\eta}{\rm 1\ MeV}\right)\left(\frac{\rm 1\ day}{\tau}\right),
\end{equation}
where $M_{\rm unstable}$ the mass of the ejected unstable strangeon nuggets (with $A<A_c$), $\Delta\eta$ is the energy released per baryon by the decay of unstable strangeon nuggets, and $\tau$ is the lifetime of the unstable strangeon nuggets.  
Therefore, if the ejected unstable strangeon nuggets constitutes about $1\%$ of the whole ejecta and have lifetime about one day, then the luminosity of the decay would be comparable to that of the observed peak of the blue component of the kilonova following GW170718~\citep{Kasliwal2017}.

The slow-fading red component could be explained by the spin-down power of the remnant strangeon star~\citep{Yu2017}.
The spin-down power evolving with time depends on the initial spin-down power $L_{\rm sd}(0)$ and the spin-down time scale $t_{\rm sd}$.
The radiation-transfer process depends on properties of the ejecta, such as the total mass $M_{\rm ej}$, the minimum and maximum velocities $v_{\rm min}$ and $v_{\rm max}$, the density distribution index $\delta$, and the opacity $\kappa$.
The bolometric light curve of a strangeon kilonova fitted to the data from~\citep{Kasliwal2017} is presented in Fig. 2, corresponding to the remnant strangeon star with $L_{\rm sd}(0)\simeq7.59\times10^{41}$ erg/s and $t_{\rm sd}=2.51\times10^5$ s, and the ejecta with $M_{\rm ej}=0.01M_{\odot}$, $v_{\rm min}=0.1c$, $v_{\rm max}=0.25c$ (with $c$ the velocity of light), $\delta=3.5$, and $\kappa=0.2$ cm$^2$ g$^{-1}$. 
More detailed demonstration of parameters in the spin-down powered kilonova is given in~\cite{Yu2017}).

   \begin{figure}[h!!!]
  \centering
   \includegraphics[width=9.0cm, angle=0]{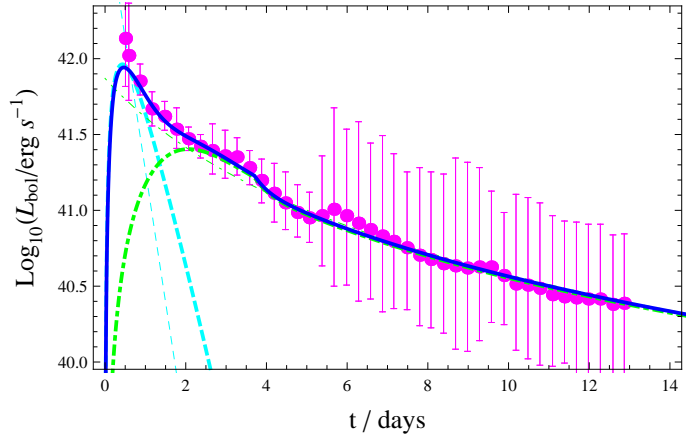}

   \begin{minipage}[]{85mm}

   \caption{Bolometric light curve of a strangeon kilonova including two energy sources, fitted to the data from~\citep{Kasliwal2017}. The thin dashed and dash-doted lines represent the heating power of decaying strangeon nuggets and strangeon star spin-down, respectively. The thick dashed and dashed-dotted lines are bolometric light curves powered by the corresponding single energy source. The thick solid line is the result of the combination of the two energy source. The ejecta has mass $M_{\rm ej}=0.01M_{\odot}$, minimal and maximum velocities $v_{\rm min}=0.1c$ and $v_{\rm max}=0.25c$, the density distribution index $\delta=3.5$, and opacity $\kappa=0.2$ cm$^2$ g$^{-1}$. The remnant strangeon star has the initial spin-down power of $L_{\rm sd}(0)=7.59\times 10^{41}$ erg/s and the spin-down timescle $t_{\rm sd}=2.51\times 10^5$ s.}\end{minipage}
   \label{fig1}
   \end{figure}

In a word, more observational tests should be necessary though the
neutron kilonova model is preferred and well testes by the the
single event of GW170817.
Developing quark kilonova as well as strangeon kilonova still
under-construction are two competition scenarios which might fit the
diverse observations, but more detailed work is surely unavoidable.

\section{The energy ejection from the thermal radiation}

We consider the merging strangeon star binaries where the mass of both stars are $1.4 M_\odot$, and the remnant strangeon stars have the mass $\sim 2.6 M_\odot$.
After the tidal disruption and merging, the remnant strangeon star at the early stages is hot and in a liquid state.
The newly formed hyper-massive strangeon star will release its internal energy by photons and neutrinos.
During this cooling process a sharp drop of temperature would lead to a phase transition from the liquid to solid state.
In the following we make a rough calculation about this cooling process, and the energy ejected to the fireball of GRB burst.
Some of our calculations are based on~\cite{YX2017} where the emission of supernova neutrinos is investigated in the strangeon star model.

In our calculation, we assume for simplicity that, inside the remnant strangeon star the number density of strangeons is uniform with $n=2.5 n_0$, where $n_0$ is the saturate nuclear matter density.
Although the number density of strangeons $n$ decreases from the center to the surface, the variation could not be significant.
Fig. 3 shows the values of $n/n_0$ as a function of distance from the center $r$, where the equation of state is from~\cite{LX2009b}, for three different values of strangeon star mass.
In the calculations, the gravity-free density of strangeon matter (i.e., surface density) is assumed to be two times of the nuclear density, for the sake of simplicity. It is evident that the density gradient would be comparably small due to the stiff state of stranegon matter.
We can see that for the case of $M\simeq 2.6 M_\odot$, $n/n_0$ only decreases from about 2.8 to 2, so we can make the approximation that the star has an uniform number density $n=2.5 n_0$.
%
   \begin{figure}[h!!!]
  \centering
   \includegraphics[width=9.0cm, angle=0]{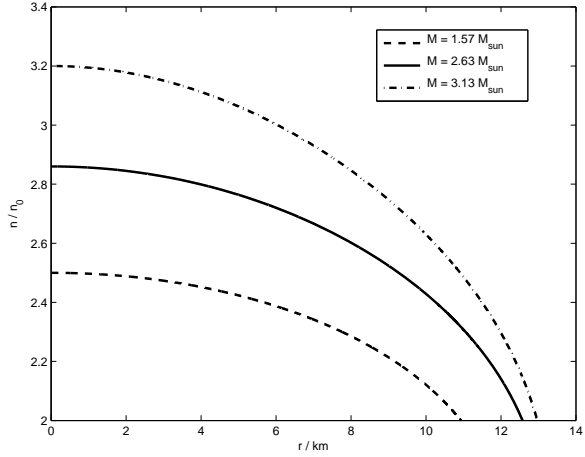}

   \begin{minipage}[]{85mm}

   \caption{Values of $n/n_0$ as a function of distance from the center $r$, where the equation of state is from~\cite{LX2009b}, for three different values of strangeon star mass with $M=1.57M_{\odot}$, $2.63M_{\odot}$ and $3.13M_{\odot}$. In the calculations, the gravity-free density of strangeon matter (i.e., surface density) is assumed to be two times of the nuclear density, for the sake of simplicity. It is evident that the density gradient would be comparably small due to the stiff state of stranegon matter.}\end{minipage}
   \label{fig1}
   \end{figure}

The internal energy of a strangeon star includes contributions from both strangeons and electrons.
The number density of electrons $n_e$  is much smaller than that of strangeons $n$, with $n_e\sim 10^{-5} n$~\citep{Alcock1986, LX2016b}.
Before solidification, the energy of electrons could be ignored~\cite{YX2017} and we only consider the energy of strangeons.
After the whole star become a solid state, however, the contribution of electrons to the heat capacity will be significant.  
The results are shown below.

\subsection{Cooling before solidification}

Before solidification, the energy of strangeon system could be estimated as
\begin{equation}
U=\frac{3}{2}nk\cdot 4\pi \int_0^R r^2 T ( r )dr,
\end{equation}
where $T( r )$ is the temperature inside the star with distance $r$ from the center, and $R$ is the radius of the star.
The newly formed strangeon star is opaque to neutrinos, so it should be non-isothermal.
Neutrinos collisions with particles inside the star, and the paths of neutrinos inside the star could be the ``random walking" process.
The relation between internal temperature $T( r )$ and the surface temperature $T_s$ could then be estimated as $T( r )\sim T_s(\frac{R-r}{l})^{1/4}$~\citep{YX2017} where $l$ is the mean free path of neutrinos.

The mean free path of neutrinos $l\simeq (n\sigma^{\prime})^{-1}$, where the cross-section $\sigma^{\prime}\simeq 0.5\times 10^{-44} {\rm cm^2} A^2(T/m_ec^2)^2$~\citep{YX2017}, and $A$ is the baryon number of each strangeon.
For simplicity we take $T$ as $T_s$, and $A=6$ which means that the number of quarks in each strangeon is 18.
The internal energy of the strangeon star could then be derived as a function of $T_s$.

The newly formed strangeon star decreases its internal energy by releasing photons and neutrinos, and the energy loss rate is
\begin{equation}
-\frac{dU}{dt}=L_{\gamma}+L_{\nu},
\end{equation}
where both $L_{\gamma}$ and $L_{\nu}$ are luminosities of thermal radiation.
The luminosity of photon radiation is $
L_{\gamma}=4\pi R^2\sigma T_s^4$, where $\sigma$ is the Stephan-Boltzmann constant.
The luminosity of neutrino radiation is $L_{\gamma}=4\pi R^2\sigma_{\nu} T_s^4$, where $\sigma_{\nu}\sim 2.3 \sigma$.

To derive the evolution of surface temperature $T_s$ with time $t$, we should know the initial temperature $T_0$ at the time $t=0$ when the remnant strangeon star is formed.
Because strangeon stars in a binary just before merger and the remnant strangeon star at the early stages should be in a liquid state, they could behave like conventional quark stars.
Simulations show that in the case of the quark star binary, the maximum temperature during the evolution is about 65 MeV if the mass of both quark stars are $1.35 M_\odot$, so we assume that in our case the surface temperature of the newly formed massive strangeon stars is $\sim 50$ MeV.
After formation, the remnant massive strangeon star will release its internal energy by photons and neutrinos.

\subsection{Cooling during and after solidification}

When the temperature drops below $\sim 1$ MeV (the melting temperature)~\citep{DX2011}, the phase transition from liquid to solid state occurs.
During this stage, the temperature will not decrease, and the latent heat $E$ would be released through thermal emission,
\begin{equation}
E=(L_{\gamma}+L_{\nu})\Delta t, \label{latent_heat}
\end{equation}
where $\Delta t$ is the time interval that the  phase transition proceeds.
The latent heat $E$ could be estimated as $E=\epsilon N$, where $N$ is the number of strangeons and $\epsilon$  MeV is the energy released by each strangeon during the phase transition.

As demonstrated before, the newborn strangeon star is non-isothermal.
The phase transition from the liquid to solid state stars at different time in different parts of the star, so the phase transition process is complicated.
To make a rough estimation about the time elapsed during the phase transition $\Delta t$, we assume that $T_s=0.7$ MeV when the internal temperature reaches 1 MeV.
Choosing the ratio of inter-strangeon potential $U_0$ to the latent heat per strangeon $\epsilon$ to be $f=0.01-0.1$, then for $U_0=100$ MeV~\citep{LX2009b}, $\epsilon=1-10$ MeV.
Due to the uncertainty about $\epsilon$, we take it as a parameter and set $\epsilon=1$ MeV, leading to $\Delta t\simeq 200$ s.
Because $\Delta t \propto \epsilon$, we can see that if $\epsilon=10$ MeV, then $\Delta t\simeq 2000$ s.
Although the value of $\Delta t$ depends on the value of $\epsilon$ and the melting temperature of strangeon stars, both of which are uncertain, we can see that $\Delta t$ could have the order of $10^3$ s, which means that the latent heat injected into the GRB fireball could explain the X-ray plateau observed in many GRBs~\citep{DX2011, Hou2017}.

When the phase transition completes, i.e., the whole star becomes solid, the internal energy of the star is then
\begin{equation}
U=\int C_v dt,
\end{equation}
where $C_v$ is the heat capacity of solid strangeon matter, including the contribution of strangeons and electrons, both of which depend on temperature.
In the Debye model~\citep{YX2011}, $C_v$ drops below $10^{35}$ erg K$^{-1}$ when temperature drops below 1 MeV, so after solidification the temperature drops very quickly.
Here we only consider the cooling from the residual thermal energy at very early stage, and do not consider heating by accretion or spin-down of the star late after its birth, because the energy ejected from the thermal radiation of the remnant strangeon star to the fireball around the star shortly after its birth could be manifested as a GRB after a prompt burst.  
If the heating processes such as the accretion heating or spin-down heating~\citep{YX2011}, the temperature of the star could be sustained at $10^6-10^7$ K in the late stages.

The cooling curve is shown in the upper panel of Fig. 4, demonstrating three stages: before solidification, during the phase transition from the liquid to solid state and after solidification, with $\epsilon=1$ MeV.
The energy ejection in the form of thermal photons from the remnant strangeon star is shown in the lower panel of Fig. 4.
The phase transition occurs when temperature drops to $\sim 1$ MeV, and the time elapsed during the phase transiton $\Delta t \simeq 10^{2-3}$ s. 
Fig.4 shows that the luminosity of the energy ejection during the phase transition is about $10^{48}$ erg, indicating that after a promote burst an X-ray plateau could follow in a timescale of $10^{2-3}$ s.

   \begin{figure}[h!!!]
  \centering
   \includegraphics[width=9.0cm, angle=0]{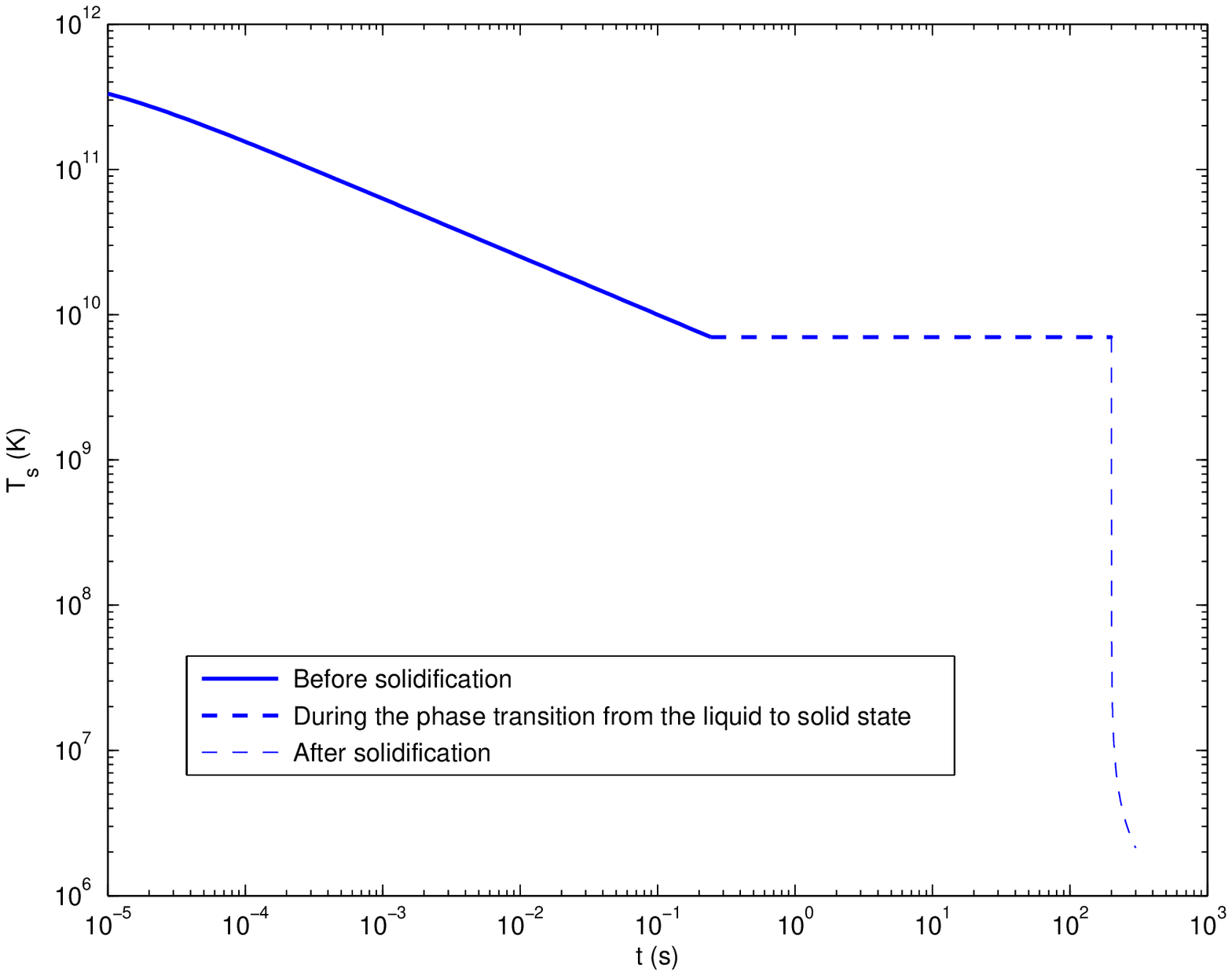}
   \includegraphics[width=9.0cm, angle=0]{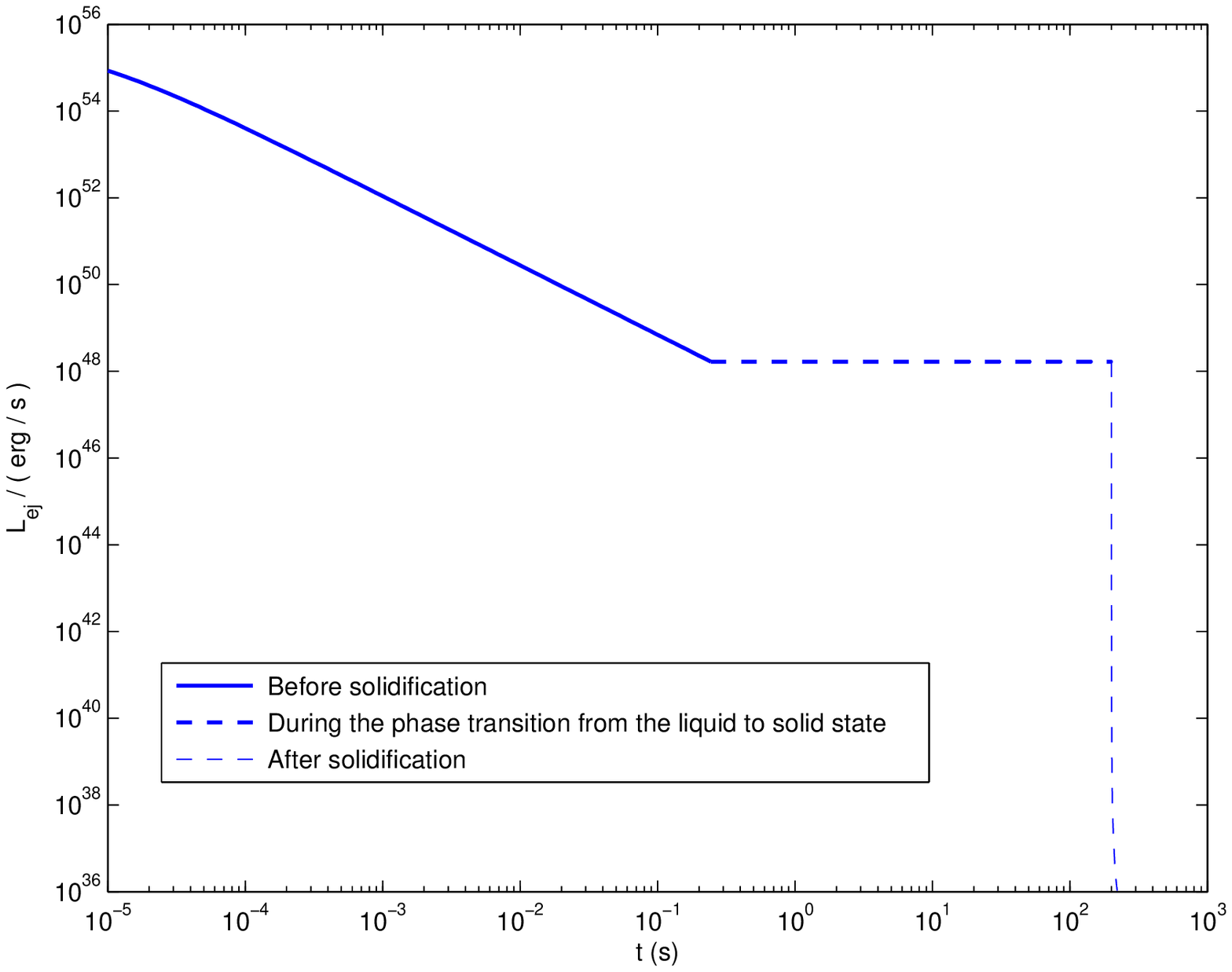}

   \begin{minipage}[]{85mm}

   \caption{Cooling curve (the upper panel) and the energy ejection from the thermal radiation of the remnant strangeon star (the lower panel), not considering the heating by accretion or spin-down of the star. There are three stages: before solidification, during the phase transition from the liquid to solid state and after solidification, with $\epsilon$ (the energy released by each strangeon during the phase transition) =1 MeV. The time elapsed during the phase transiton $\Delta t \simeq 200$ s. If $\epsilon=10$ MeV, then $\Delta t \simeq 2000$ s. The The phase transition occurs when temperature drops to $\sim 1$ MeV. This means that after a promote burst, an X-ray plateau could follow in a timescale of $10^{2-3}$ s.}\end{minipage}
   \label{fig2}
   \end{figure}

\section{Summary and Discussions}

Pulsar-like compact stars could actually be the so-called ``strangeon stars''.
The merger of a strangeon star binary composed of two strangeon stars with mass $\sim 1.4 M_\odot$ could result in the formation of a hyper-massive strangeon star with mass $\sim 2.6 M_\odot$, accompanied by bursts of gravitational waves and electromagnetic radiation.
In this paper we discuss the corresponding electromagnetic behavior.
From the constraints on tidal polarizability by GW170817, we find that the strangeon star model is more favored than the neutron star model.
The ``strangeon kilonova'' scenario is introduced, which could be powered by the decay of ejected strangeon nuggets and the spin-down of the remnant strangeon star.
The energy ejection from the thermal radiation of the remnant strangeon star is also calculated, which shows that an X-ray plateau could follow in a timescale of $10^{2-3}$ s.
Our result could be tested by observations, i.g., by HXMT and future eXTP.

In the strangeon kilonova scenario, we show that the bolometric light curve of the kilonova following GW 170817 could be explained by combining two energy source including the decay of ejected strangeon nuggets and the spin-down of the remnant strangeon star. 
It is worth mentioning that there are two other possibilities.
On one hand, the nucleosynthesis of protons and neutrons from the decaying strangeon nuggets could produce radio-active heavy elements whose decay could also contribute to the slow-fading red component in the kilonova following GW170817.
On the other hand, if all the ejected strangeon nuggets are stable, only the spin-down of the remnant strangeon star could also power the bolometric light curve.

More observational tests should be necessary though the
neutron kilonova model is preferred and well testes by the the
single event of GW170817.
Developing quark kilonova as well as strangeon kilonova still
under-construction are two competition scenarios which might fit the
diverse observations, but more detailed work is surely unavoidable.
Certainly, our results could be tested also by further observational synergies between gravitational wave detectors (e.g., aLIGO) and X-ray telescopes (e.g., Chinese HXMT and eXTP), and especially if the detected gravitational wave form is checked by peculiar equation of state provided by the numerical relativistical simulation.

The coalescence of neutron stars in binaries is taken as the origin of the gamma-ray bursts (GRBs)~\citep{Paczynski1986}.
The fireballs in GRBs associated supernovae could be explained if strangeon stars are formed in core-collapse supernovae~\citep{CYX2007}, then it is possible that merging strangeon stars can give rise to short GRBs, although no calculation or simulation has been performed yet.
Interestingly, strange nuggets in the torus (or disk) could behave like dust if each of them contains high enough baryon number ($\sim 10^{20-30}$).
These strange nuggets might absorb the X-ray from the newly formed strangeon star and could reradiate in infrared wavelengths, which would be tested by observations.

\normalem
\begin{acknowledgements}
We would like to thank the anonymous referee for the valuable comments.
This work is supported by the National Key R\&D Program of China (No. 2017YFA0402602), the West Light Foundation (XBBS-2014-23), and the National Natural Science Foundation of China (11203018,11673002, U1531243).
\end{acknowledgements}

\label{lastpage}

\end{document}